\documentclass[12pt]{article}

\usepackage{epsfig}
\usepackage{graphicx}
\usepackage{color}

\newcommand {\hAT} {0.382}
\newcommand {\qmin} {0.21}

\newcommand {\bi} {\bibitem}
\newcommand {\be} {\begin{equation}}
\newcommand {\beq} {\begin{eqnarray} \nonumber }
\newcommand {\ee} {\end{equation}}

\newcommand{\bq}{\begin{eqnarray}}
\newcommand{\eq}{\end{eqnarray}}

\newcommand{\bc}{\begin{center}}
\newcommand{\ec}{\end{center}}

\begin{document}
\title{
Magnetic field chaos in the SK Model
}
\author{Alain Billoire and Barbara Coluzzi  }

\maketitle

\begin{center}

{\em Service de Physique Th\'eorique} CEA-Saclay \\
Orme des Merisiers - 91191 Gif sur Yvette France


\end{center}

\vspace{2cm}

\begin{abstract}
\noindent 
We  study the Sherrington--Kirkpatrick model, both above and
below the De Almeida Thouless line, by using a modified version of the
Parallel Tempering algorithm in which the system is allowed to move
between different values of the magnetic field~$h$. The behavior of
the probability distribution of the overlap between two replicas at
different values of the magnetic field $h_0$ and $h_1$ gives clear
evidence for the presence of magnetic field chaos already for moderate
system sizes, in contrast to the case of temperature chaos, which is not
visible on system sizes that can currently be thermalized. 
\end{abstract}

\newpage

\begin{section}{Introduction}
\noindent
The Sherrington--Kirkpatrick (SK) model was introduced quite a long time
ago \cite{ShKi} as a mean field model for spin glasses. Its proposed analytical
solution \cite{Pa} displays intriguing features such as an infinite 
number of pure states in the glassy phase, described by an order parameter
which is the non trivial probability distribution of the overlap between  two states, 
$P(q)$. After more than twenty years this solution is still the subject
of works aiming at establishing it in full mathematical rigor \cite{GuTo,Gu}, 
whereas long standing open issues concern the study of the corrections to the 
mean field approximation below the upper critical dimension \cite{DeDo} and 
the very applicability of the mean field picture to short range realistic 
spin glasses \cite{MaPaRiRuZu}.

An interesting question concerns the way in which the states reorganize
themselves when the system is subjected to a small perturbation $\delta p$
of an external parameter, in particular the temperature $T$ or the
magnetic field $h$. There is the intriguing possibility of  $p$ chaos, 
namely that states at $p$ and states at $p+\delta p$ are as different as 
possible in the thermodynamic limit.

The possible presence of  temperature chaos in the SK and related
models is an old subject of investigations \cite{Ko}-\cite{FrNe} that
recently received a lot of attention both analytically and numerically 
\cite{BiMa}-\cite{SaMa}. From a very recent analytical computation 
\cite{CrRi} it turns out to be present but to be of the ninth order in 
perturbation theory, a very weak effect, extremely difficult to be 
numerically observed on the system sizes one is currently able to thermalize. 

The aim of this paper is to investigate the appearance of chaos with
increasing system sizes (a question that cannot be addressed by
existing analytical techniques, that are restricted to the asymptotic $N \to \infty$
regime), in a case where chaos is strong, namely the case of magnetic field chaos.
The presence of magnetic field chaos was
predicted already twenty years ago \cite{Pa1} (see also
\cite{Ko,FrNe}). From the numerical
point of view, it was observed in a previous work \cite{Ri} from a
study of the behavior of the second moment of the probability
distribution of the overlap $P_{h_0,h_1}(q)$ between replicas at
$h_0=0$ and $h_1\neq0$.  This pioneering paper can however be
criticized since many data points are on the wrong side of the 
De Almeida Thouless (AT)
line \cite{AlTh}. We will revisit the problem by looking (in the SG phase) at
the distribution $P_{h_0,h_1}(q)$ itself, a quantity whose
interpretation is simpler than the moments.

More in detail, in terms of the probability distribution of the
overlap between two replicas at different values of the external
parameter $h_0$ and $h_1=h_0+\delta h$, chaos has a very clean
signature. Taking for simplicity the case $h_0=0$, for small volumes
$P_{0,\delta h}(q)$ has two peaks, and is very similar to $P_{0,0}(q)$
on the same volume.  As the volume grows, a peak develops around the
minimal value of the overlap $q_m=0$, in such a way that for very
large volumes $P_{0,0}(q) \approx \delta(q)$.  In the temperature chaos
case, this chaotic peak is hardly visible with current computers and
algorithms. Our aim is to determine if and how this ``chaos peak'' scenario takes
place in the case of $h$ chaos, that is believed to be much stronger
than $T$ chaos.

To this aim, we perform  numerical simulations of the SK model
at $T=0.6T_c$, both above and below the AT line, 
by using a modified version of the Parallel Tempering algorithm 
\cite{TeReOrWh,Ma} in which the system is allowed to move between different 
$h$ values at fixed temperature.

\end{section}

\begin{section}{Model and Observables}
\noindent
The Sherrington--Kirkpatrick spin glass model \cite{MePaVi,BiYo} is described 
by the Hamiltonian
\be
{\cal H}_{J}=\sum_{1\le i < j \le N} J_{ij} \sigma_i \sigma_j
-h\sum_{1\le i \le N} \sigma_i,
\ee
where $\sigma_i=\pm 1$ are Ising spins, the sum runs over all pairs of
spins and $J_{ij}$ are quenched identically distributed independent
random variables with mean value $\overline{J_{ij}}=0$ and variance
$1/N$. We take $J_{ij}=\pm N^{-1/2}$.

In order to measure the probability distribution of the overlap $P(q)$
one usually considers two independent replicas  $\{ \sigma_i \}$ and 
$\{ \tau_i \}$ evolving contemporaneously and independently (at the same
temperature and at the same value of the magnetic field):
\begin{eqnarray}
{\cal Q}&=&{1 \over N} \sum_{i=1}^N \sigma_{i} \tau_{i} \\
P(q)&\equiv&\overline{P_J(q)}\equiv
\overline{ \langle \delta(q-{\cal Q})\rangle},
\end{eqnarray}
where the thermal average $\langle \cdot \rangle$ corresponds to
the average over Monte Carlo time in the simulation whereas 
$\overline{(\cdot )}$
stands for the average over the $J_{ij}$ realizations.  This is the
order parameter in the glassy phase, which in the thermodynamic limit
behaves as
\begin{equation}
P(q)=\left \{
\begin{array}{lcl}
\delta(q-q_{EA}) & \hspace{.3cm} & |h|>h_{AT}(T) \\
x_m \delta(q-q_m) +\tilde{P}(q)+ x_M \delta(q-q_{EA}) 
& \hspace{.3cm} & 0<|h|<h_{AT}(T) \\
\frac{1}{2} \left [\tilde{P}(q)+\tilde{P}(-q) \right ] +
{x_M \over 2} \left [ \delta (q-q_{EA}) + \delta (q+q_{EA}) \right ]
& \hspace{.3cm}& h=0, T<T_c
\end{array}
\right .
\end{equation}
where $h_{AT}(T)$ is the critical value of the magnetic field signaling
the AT line, with $h_{AT}(T) \sim (4/3)^{1/2}(T_c-T)^{3/2}$ for 
$T \rightarrow T_c^-$ ($T_c=1$
in this model) \cite{AlTh}. 
In the glassy phase, the stable solution corresponds
to a full replica symmetry breaking (FRSB), i.e. to a non-trivial
$P(q)$ with a continuous distribution $\tilde{P}(q)$ between two 
$\delta$-functions at values $q_{EA}$ and $q_m$ respectively. For 
$T \rightarrow T_c^-$ one finds that
$x_m \propto q_m \propto h^{2/3}$, $(q_{EA}-q_m) 
\propto (x_M-x_m) \propto (h_{AT}(T) -h)$.
Note that at $h=0$ the function $P(q)$ is
symmetric, reflecting the symmetry of the system for 
 $\{ \sigma_i \} \rightarrow \{- \sigma_i \}$, and the $\delta$-function
in $q_m$ disappears.

The interesting quantity to study when looking for chaos is the probability
distribution of the overlap between two replicas which evolve at different values
of the magnetic field, $h_0$ and $h_1=h_0+\delta h$, 
definable as
\be
P_{h_0,h_1}(q)=\overline{ \langle \delta(q-{\cal Q}_{h_0,h_1})\rangle}.
\ee
It is expected to become a $\delta$-function in the thermodynamic limit,
where in presence of chaos states are as different as possible and accordingly
their mutual overlap approach the minimum possible value, i.e. $q_m(h_0)$ 
(which is zero for $h_0=0$). This happens certainly in the 
$N\rightarrow \infty$ limit as soon as the condition $(h_1-h_0)^2N>>1$ is 
verified \cite{Pa1}.

In finite dimensions, one can define the
overlap correlation function 
$C_{h_0,h_1}(|\bf{r}_i-\bf{r}_j|)=
\overline{\langle\sigma_i\sigma_j \rangle\langle\tau_i \tau_j \rangle}$ 
which decays exponentially with a correlation length that was evaluated
(in $d>8$, i.e. above the upper critical dimension of the model)
\cite{Ko} to be $\xi_{h_0=0,h_1}\propto {h_1}^{-2/3}$ and 
$\xi_{h_0\neq0,h_1=h_0+\delta h}\propto {h_0}^{-1/6}(\delta h)^{-1/2}$ 
respectively.

Adimensional ratios of momenta, such as
\begin{eqnarray}
A^{2n}(h_0,h_1,T)&=&\frac{\overline{\langle \left ( q-\overline
{\langle q \rangle}_{h_0,h_1} 
\right )^{2n} \rangle}_{h_0,h_1}}
{\sqrt{\overline{\langle \left ( q-\overline{\langle q 
\rangle}_{h_0,h_0} \right )^{2n} 
\rangle}_{h_0,h_0}\hspace{.1in}\overline{\langle \left ( q-
\overline{\langle q \rangle}_{h_1,h_1} \right )^{2n} \rangle}_{h_1,h_1}}}, \\
B^{2n}(h_0,h_1,T)&=&\frac{\overline{\langle \left ( q-
\overline{\langle q \rangle}_{h_0,h_1} 
\right )^{2n} \rangle}_{h_0,h_1}}
{\overline{\langle \left ( q-\overline{\langle q 
\rangle}_{h_0,h_0} \right )^{2n} 
\rangle}_{h_0,h_0}},
\end{eqnarray}
have been introduced in order to look for chaos in reference \cite{Ri}, where it
was argued that they should scale as $\tilde{f}(N h_1^{8/3})$ for
$h_0=0$ and approach zero for $N \rightarrow \infty$, namely that
there is magnetic field chaos.

The finite size corrections to the asymptotic behavior of
$P_{h_0,h_1}(q)$ were computed in \cite{FrNe} by considering two
replicas, at different values of the magnetic field, constrained to have
a fixed overlap $q$. The constraint causes a free energy excess for
$q\ne q_{m}$ given by  $\Delta f=f(q=q_{m}+\delta q)-f(q=q_{m})$, with
\be
\Delta f = \left \{
\begin{array}{lcl}
\left(\frac{\textstyle 2187}{\textstyle 32}\right)^{1/3}
\frac{\textstyle \delta q^2 h_1^{8/3}}{\textstyle q_{EA}} & 
\hspace{.3in} & h_0=0 \\
&&\\
\frac{\textstyle \delta q^2 h_0 \delta h}{\textstyle \sqrt{2}} & 
\hspace{.3in} & h_0\neq 0,
\delta  h=|h_1-h_0 | << h_0 \\
\end{array}
\right.
\label{pqh1h2}
\ee
Correspondingly one has $P_{h_0,h_1}(q)\propto \exp({-N\Delta f})$, i.e.
a Gaussian with variance  $\overline{\langle q^2 \rangle_{0,h_1}}\propto (N h_1^{8/3})^{-1}$
for $h_0=0$, in agreement with the above scaling law.

\end{section}

\section{Parallel Tempering in Magnetic Field}
The Parallel Tempering (PT) or Multiple Markov Chain Method 
is a widely used numerical algorithm particularly efficient for 
simulating (some) systems with a corrugated free energy landscape. The basic idea is 
that the system at equilibrium, instead of being trapped in a single low 
temperature valley is allowed to move at higher 
temperatures where the landscape is trivial and to return at low $T$ in a 
different valley. This can be achieved by considering $n$ replicas of 
the system, each one at a different temperature in a given set (of temperatures), and by 
allowing exchanges of temperatures between nearest neighbor replicas with 
the usual Monte Carlo probability.

Here we consider a set of replicas at different values of the magnetic
field, both above and below the AT line, allowing exchange of $h$ values
between nearest neighbor replicas with the appropriate probability
\be
P\left(\left \{h_1,\{\sigma^1\}; h_2,\{\sigma^2\}\right\} 
\rightarrow \left\{h_2,\{\sigma^1\};h_1,\{\sigma^2\}\right\} \right)=
\left\{
\begin{array}{lcl}
1 &\hspace{.3in}& \Delta H_{tot} > 0 \\
e^{\beta  \Delta H_{tot} } &\hspace{.3in}& \Delta H_{tot} < 0
\end{array}
\right.
\end{equation}
where
\be
H_{tot}=\sum_{a=1}^{n} \sum_{1\le i<j \le N} J_{ij} \sigma_i^a \sigma_j^a
-h_{\pi(a)} \sum_{1 \le i \le N}\sigma_i^a
\ee
and therefore
\be
\Delta H_{tot}=-(h_1-h_2) \left ( \sum_{i=1}^N\sigma_i^1-\sum_{i=1}^N\sigma_i^2
\right)
\ee
In principle this  h-PT method should be efficient for thermalization like the usual
T-PT method, since  the landscape is trivial above the AT line. It is moreover a well
suited method for the kind of numerical study we are interested in, since one 
can easily measure $P_{h_0,h_1}(q)$ by considering two or more independent 
sets of replicas. However, as we are going to discuss in detail, we find that 
its efficiency rapidly decreases when simulating large system sizes.

We studied the cases $N=64,256$ and 1024, taking a set of $n=49$ equally 
spaced magnetic field values between $h=\pm |h_{max}|=\pm 0.6$ at the 
temperature $T=0.6$, where the AT line occurs at the critical value 
$h_{AT}(T=0.6)\simeq\hAT$ \cite{CrRi1}. 

We alternate one sweep of each replica with the usual Metropolis algorithm
and one sweep with the PT algorithm.

\begin{figure}[htbp]
\begin{center}
\leavevmode
\epsfig{figure=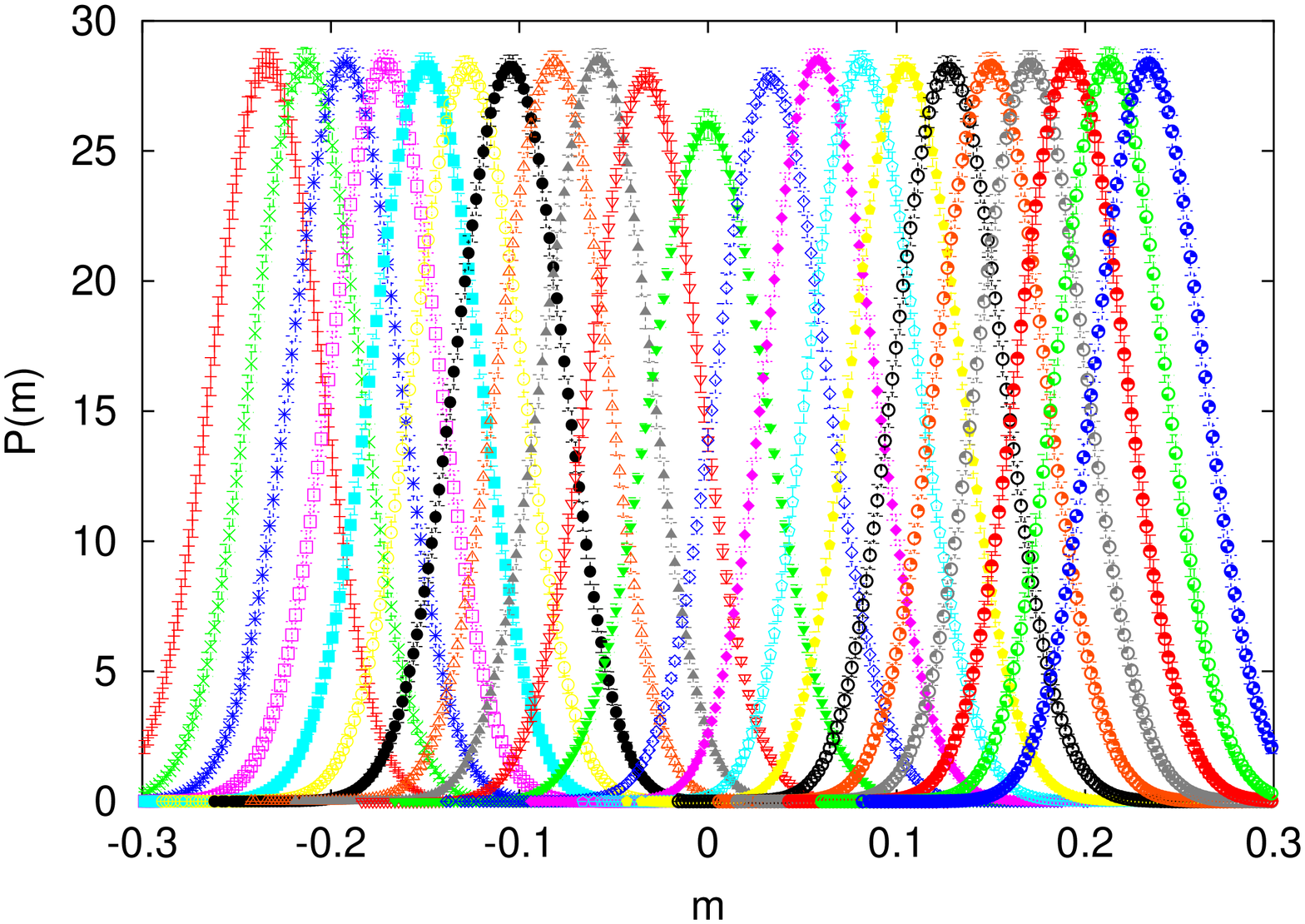,angle=270,width=8cm}
\epsfig{figure=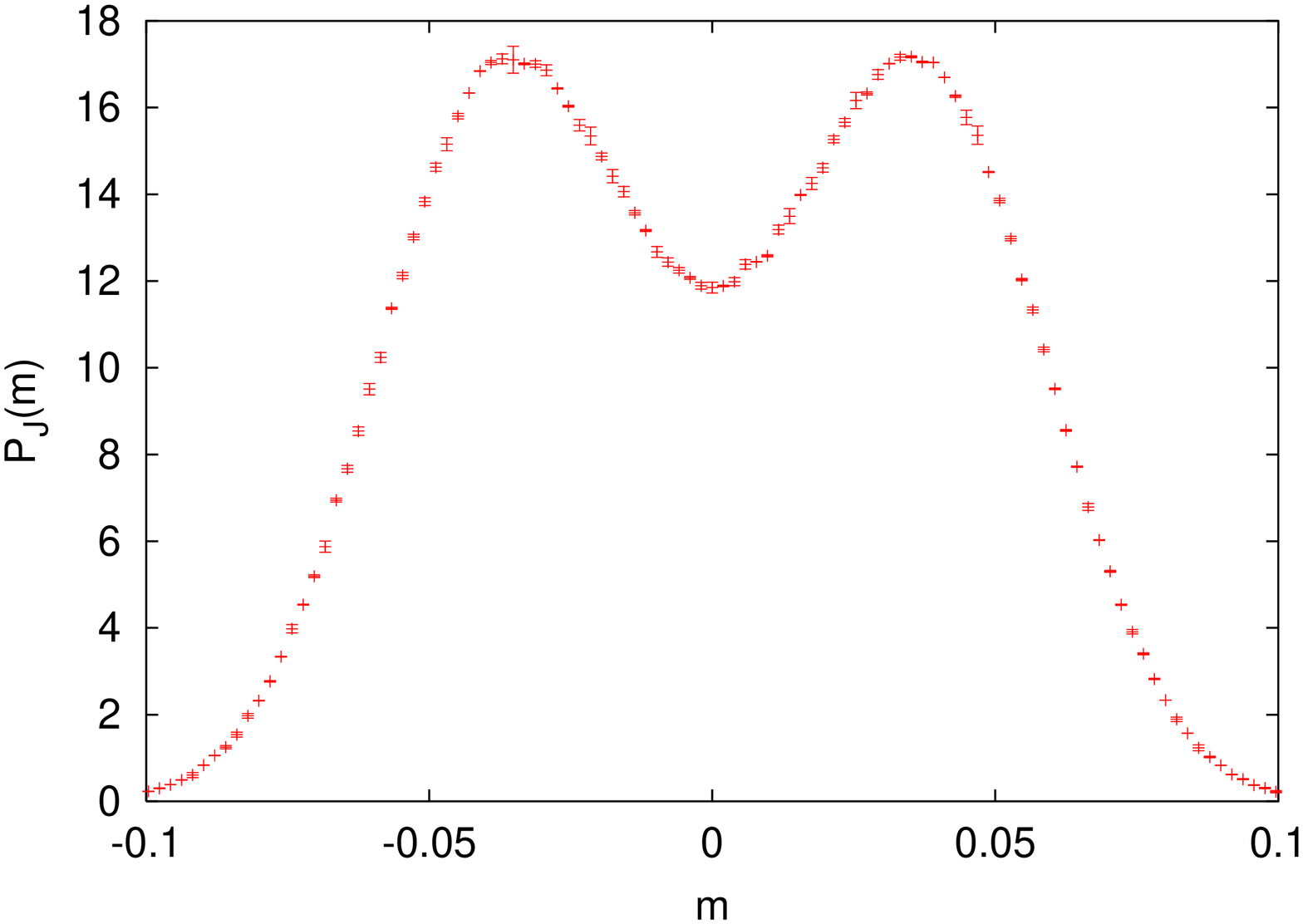,angle=270,width=8cm}
\caption{On the left we plot the disorder averaged probability distribution of 
the magnetization $P(m)$ at the 21 different central $h$ values of the set 
(i.e. from $h=-0.25$ to $h=0.25$) for the largest considered system sizes
$N=1024$. On the right we present $P_J(m)$ at $h=0$ for a two-peak 
sample  for $N=1024$ again. In this last case the errors are roughly
evaluated as the difference between the values measured in the second quarter 
and in the second half of the run.}
\end{center}
\end{figure}

The probability for two replicas to exchange their magnetic fields is
related to the overlap between the corresponding histograms of the
magnetization $P(m)$ that we check to be large enough (see [Fig. 1])
even for $N=1024$.  However some single sample $P_J(m)$ display two
peaks at $\pm m_0 \neq 0$ when $h=0$ (see [Fig 1]). As a result, replicas can
separate into two distinct subsets, one evolving in the phase space
with positive and the other with negative magnetic field values (the
probability for a replica which arrive at $h=0$ with $m \simeq -m_0<0$
to move at a positive $\delta h$ value is of order $\exp(-\beta m_0
\delta h N)$, much smaller than the usual $\exp(-\beta
\chi {\delta h}^2 N)$). This happens for some $N=1024$ samples. In order to avoid
such a problem, we add a new possible global movement, allowing
a replica at $h=0$ to reverse the sign of all its spins with probability
1/2.

Each run is divided into two equal parts and we check thermalization by 
comparing the data obtained in the second part with the ones of the second 
quarter, looking in particular at the behavior of $P_{h_0,h_1}(q)$.
We perform 50.000+50.000, 100.000+100.000 and 300.000+300.000 h-PT steps for
$N=64$, 256 and 1024 respectively. In the $N=1024$ case we also performed 
independent runs with temperature PT for 64 disorder samples at $h=0$ and 
$h=0.3$, obtaining indistinguishable results for $P(q)$.

We simulated four sets of replicas evolving contemporaneously and
independently (i.e. $49 \times 4=196$ replicas). Data are averaged
over 256 disorder configurations for each system size, and statistical
errors are evaluated from sample-to-sample fluctuations by using the
Jack-knife method. The program was multi-spin coded with 64 different
sites of the system in the same computer word and the whole
simulations took about 5500 CPU hours (in the largest part used for
$N=1024$), i.e. about one week when running over 32 processors on the
COMPAQ SC270 (the program can be easily parallelized by running
different samples over different processors).

In the $N=64$ and $256$ cases the algorithm works quite nicely, as can
be seen from the number of tunnelings, namely the number of times that
each replica moves from one extrema of the set (of $h$ values) to the
other and back, which is about ${\cal N}= 15 \div 20$ (in the second
half of the run).  On the other hand, already for $N=1024$, despite
the 300.000 PT steps of the second part of the run, this number drops
to ${\cal N}=5\div 6 $ and in nearly 1/4 of the samples there is at
least one replica which is unable to go from $h_{max}$ to $h_{min}$
and back in the whole considered interval (in a few cases most
replicas never did it).

In the case of temperature PT, the corresponding (average) number of
tunnelings are $3780$, $1590$ and $455$ respectively (in runs of
400.000 steps starting from equilibrium configurations, with a set of
38 temperatures between $T_{max}=1.325$ and $T_{min}=0.4$ at $h=0.3$
for $N=64, 256$, and $1024$). Clearly the number of tunnelings
decreases much faster with system size in the h-PT case. This is
presumably linked to the early appearance of magnetic field chaos.  As
we will discuss in detail in the next section, $P_{h_0,h_1}(q)$ starts
to approach a $\delta$-function, i.e. its thermodynamic limit
behavior, already for magnetic field differences of order 0.15, for
$N=1024$. This means that the corresponding phase spaces are $very$
different and that an algorithm based on global movements between
different values of $h$ can not work well.  Similarly the efficiency
of temperature PT should drop down for extremely large systems due to
temperature chaos finally coming out.

The bottom line is that $N=1024$ is the largest size we are able to
efficiently thermalize at $T=0.6$ by using the h-PT algorithm, to be
compared with the four time larger $N=4096$ that can be thermalized
down to $T=0.4$ with the  temperature PT algorithm at zero magnetic field.

\begin{section}{Results and Discussion}
\begin{subsection}{On the finite size corrections to the $P(q)$.} 
\noindent
The function $P(q)$ is shown in [Fig. 2] for $h=0.0,0.1,0.2$ and
0.3. At $h=0.0$ it agrees nicely with the expected behavior, whereas
it is strongly affected by finite size effects for non-zero magnetic
field. This is in qualitative agreement with the theoretical finding
\cite{FrPaVi} that the finite size corrections of $P(q)$ 
are an order of magnitude larger in the $q<q_m$ region than in the
$q>q_{EA}$ region, namely,
\be
\frac{\ln\left ( P(q)\right )}{N}=\left\{
\begin{array}{lcl}
-\lambda_m(q_m-q)^3 &\hspace{.3in}& q<<q_m\\
-\lambda_{EA}(q-q_{EA})^3 &\hspace{.3in}& q>>q_{EA}\\
\end{array}
\right.
\label{pqn}
\ee
with $\lambda_m<<\lambda_{EA}$. The behavior for $q>q_{EA}$ was  tested (for $h=0$) in
\cite{PaRiSl} and \cite{BiFrMa}.

\begin{figure}[htbp]
\begin{center}
\leavevmode
\epsfig{figure=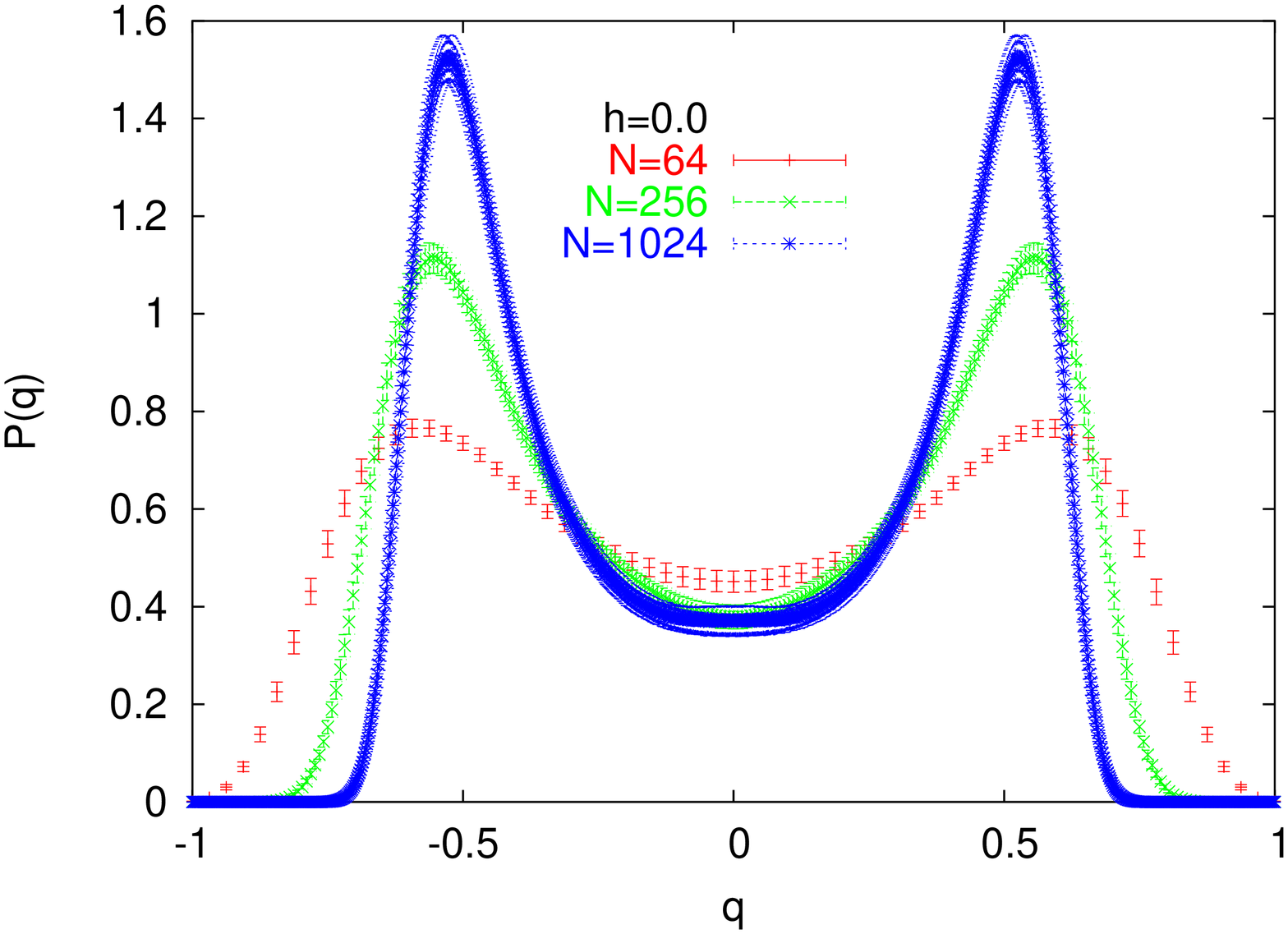,angle=270,width=8cm}
\epsfig{figure=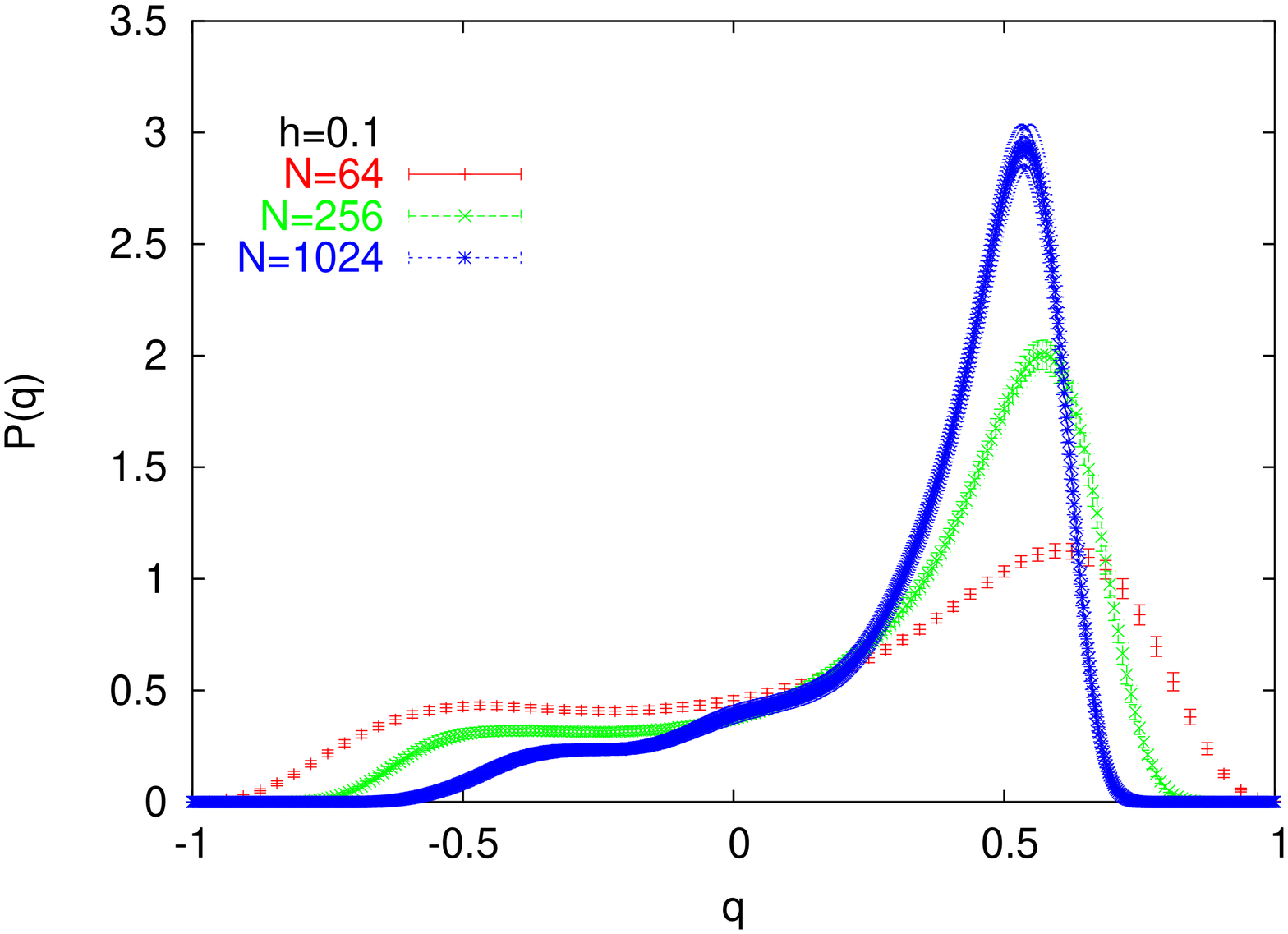,angle=270,width=8cm}
\epsfig{figure=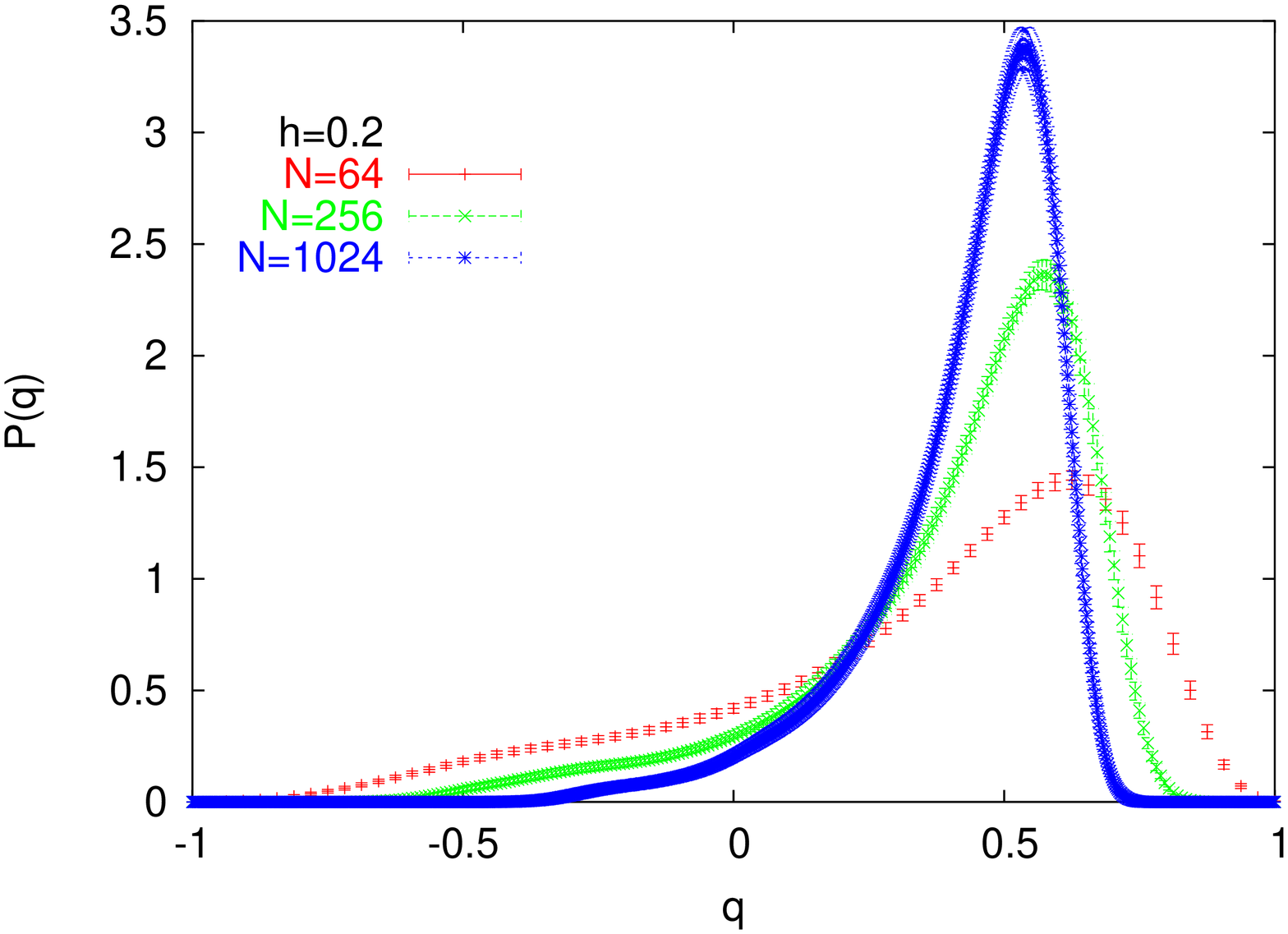,angle=270,width=8cm}
\epsfig{figure=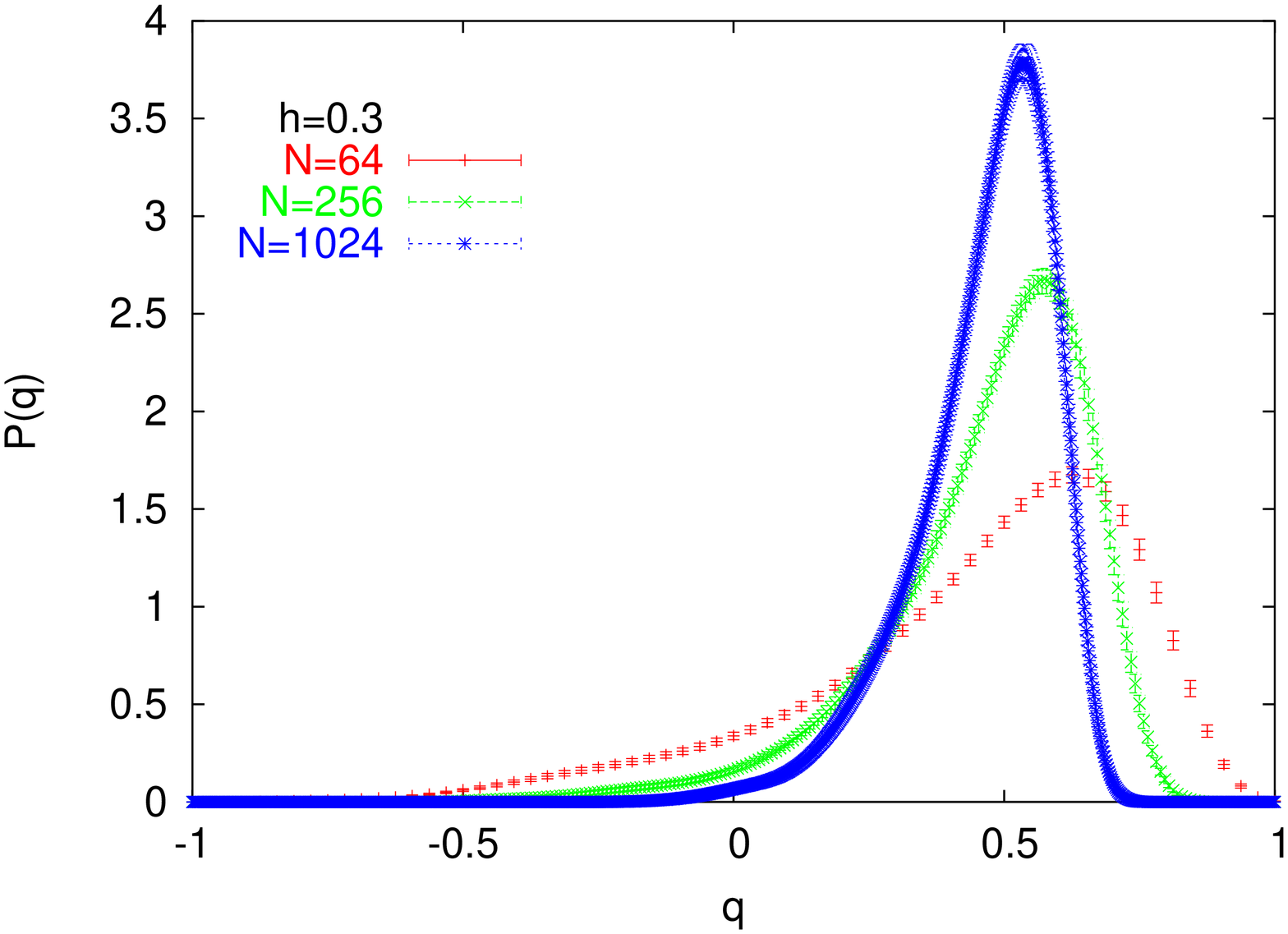,angle=270,width=8cm}
\caption{The probability distribution of the overlap $P(q)$ between two replicas 
evolving at $h=0.0, 0.1, 0.2$ and 0.3 respectively, for the considered
system sizes.}
\end{center}
\end{figure}
Here we find that for the considered sizes $P(q)$ has a visible tail in the
$q<0$ region for $h$ as large as $0.3$. Moreover the peak that should
correspond to the thermodynamic limit $\delta(q-q_m)$ is not visible
and the expected $\exp({-N\lambda_m (q_m-q)^3})$ behavior is swamped by
the reminiscence of the $q=-q_{EA}$ peak, still clearly visible at
$h=0.1$. For increasing magnetic fields the weight of the
$q=q_m$ peak should increase (and the reminiscence of the $q=-q_{EA}$
peak fade away) but $q_m$ approaches $q_{EA}$ making difficult to
distinguish between the two peaks. This kind of strong finite size
effects in magnetic field were already observed in finite dimensional
spin glasses \cite{CiPaRiRu,PiRi}. Larger system sizes and / or lower temperatures
would be needed in order to see the correct large volume behavior.

On the other hand, we note that in our data  $q_{EA}$ (defined as the
location of the maximum of $P(q)$) is practically independent of the
field, as predicted by Parisi theory (in the infinite volume limit).
We obtain $q_{EA} \simeq 0.53$ for $N=1024$, where a recent analytical
computation \cite{CrRi1} gives the asymptotic value $q_{EA} \simeq
0.505$ (independently of $h$).

\end{subsection}

\begin{subsection}{On magnetic field chaos}
\noindent
In order to find evidences for magnetic field chaos we analyze the
behavior of $P_{h_0,h_1}(q)$.
We first consider the case $h_0=0.0$ (then $P_{0,h_1}(q)$ is still
symmetric for $q\rightarrow -q$) and let $h_1$take the values 0.1,
0.15, 0.2 and 0.3 (see [Fig. 3]). Already for $h_1=0.15$ we find clear
evidences for a chaotic behavior when looking at the $N=1024$ data.  {This
is very different from the situation one finds when looking for
temperature chaos \cite{BiMa} where $P_{T_0,T_1}(q\approx 0)$ does not
show a clear peak corresponding to the thermodynamic limit $\delta(q)$
for $(T_1-T_0)$ as large as 0.2 and sizes as large as $N=4096$.} 
It is remarkable
that the appearance of magnetic field chaos with increasing system
sizes is a very sudden phenomenon: chaos is
elusive for $N=256$ and  blatant for $N=1024$.

\begin{figure}[htbp]
\begin{center}
\leavevmode
\epsfig{figure=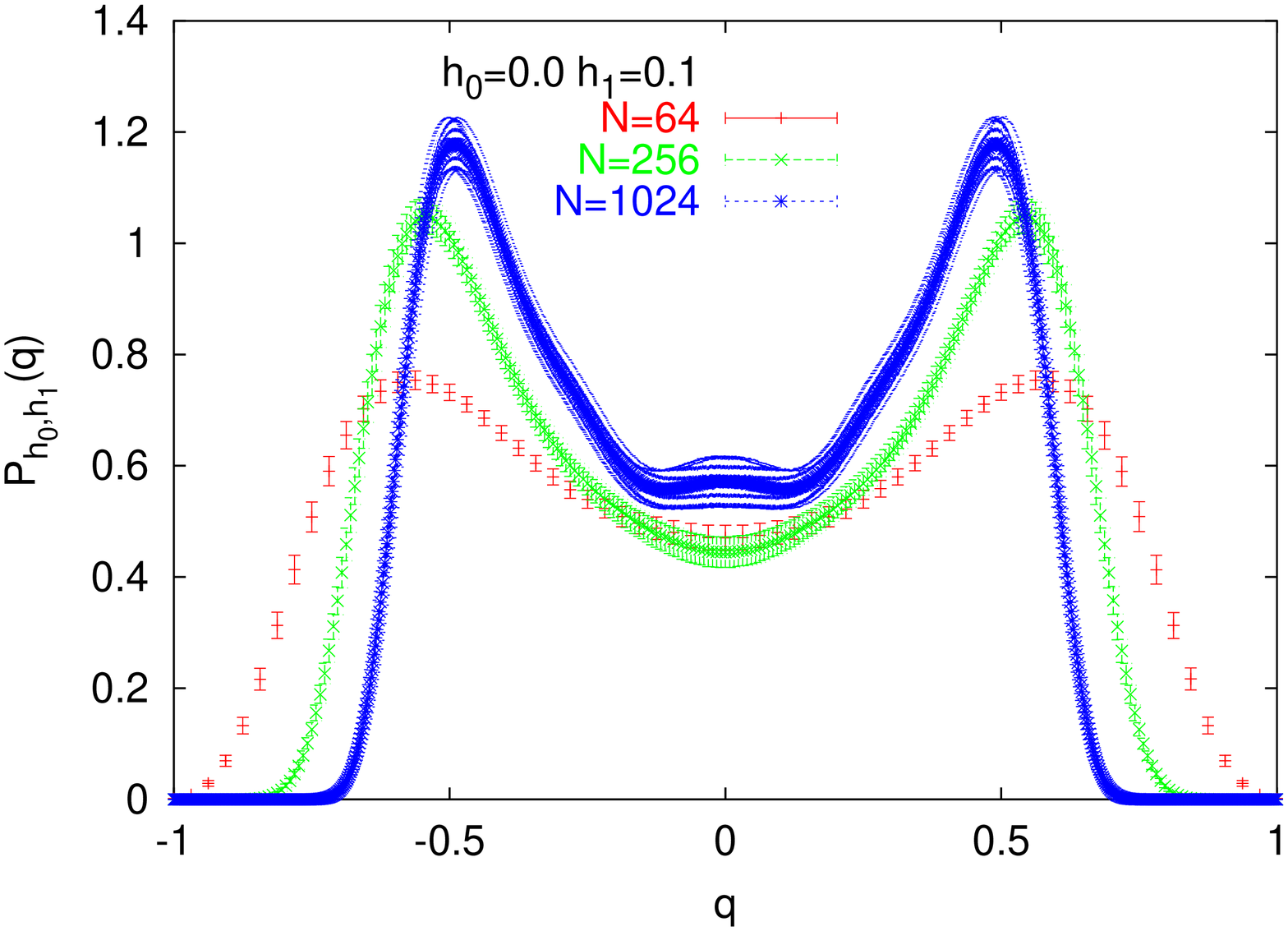,angle=270,width=8cm}
\epsfig{figure=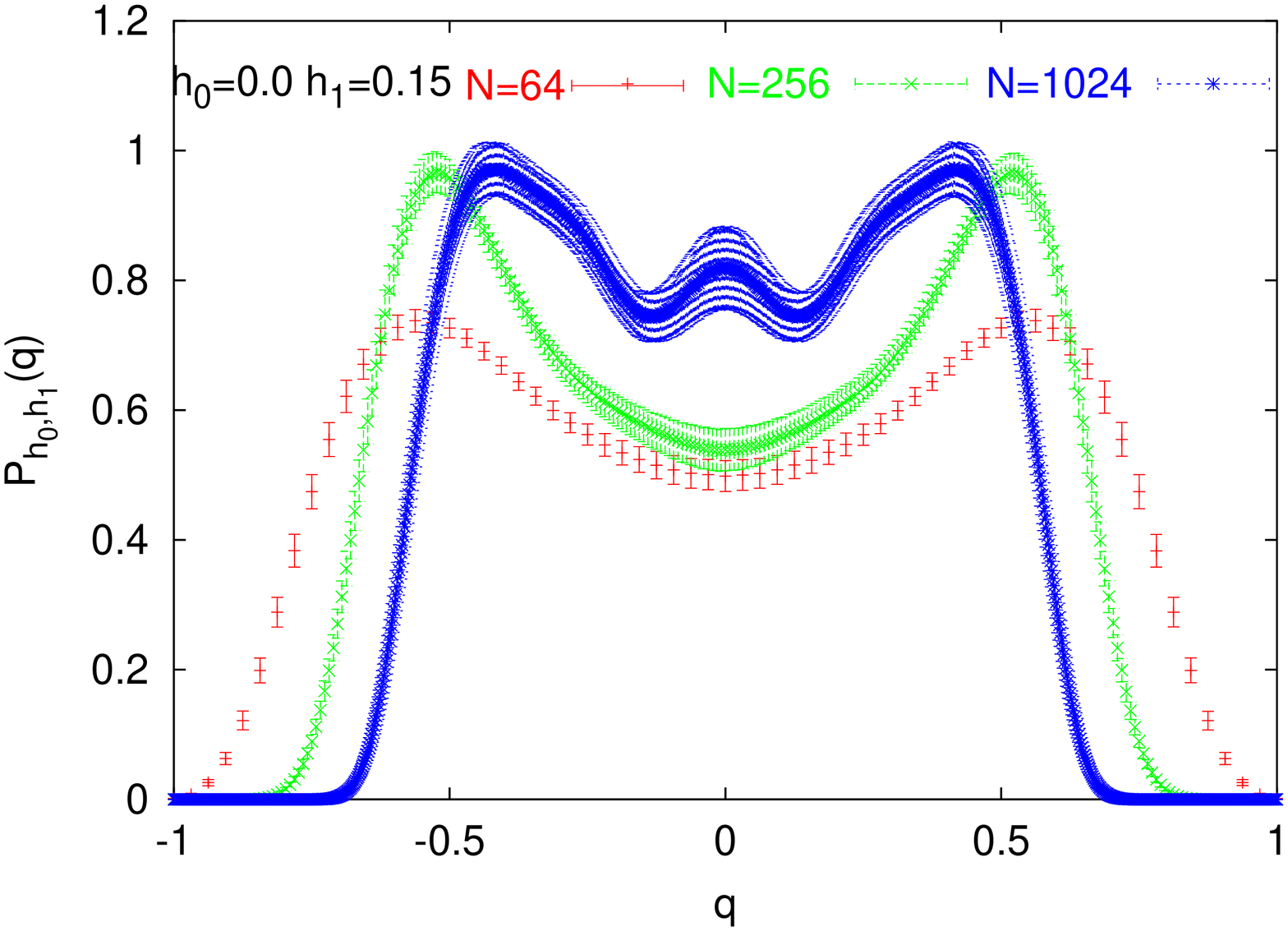,angle=270,width=8cm}
\epsfig{figure=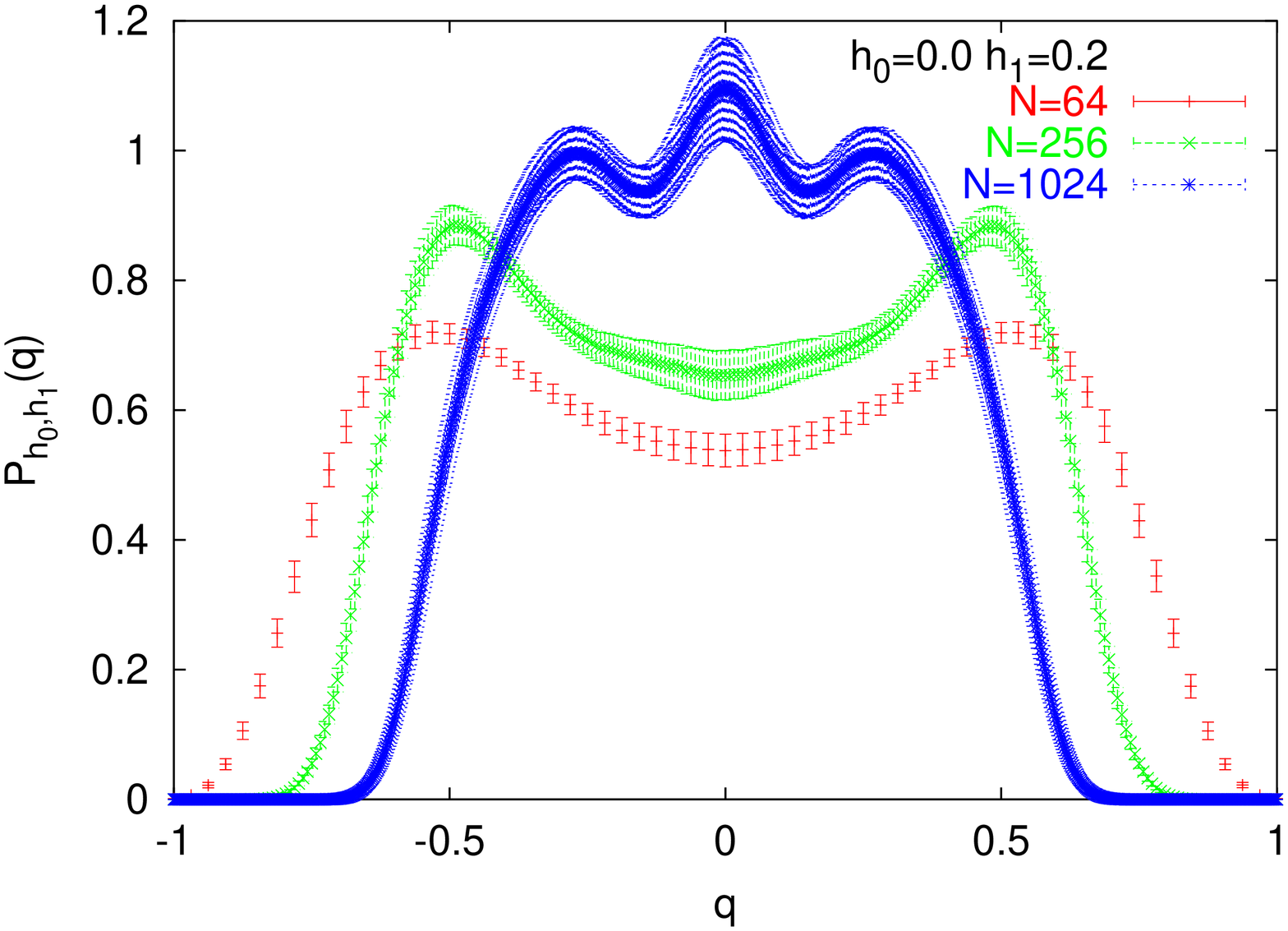,angle=270,width=8cm}
\epsfig{figure=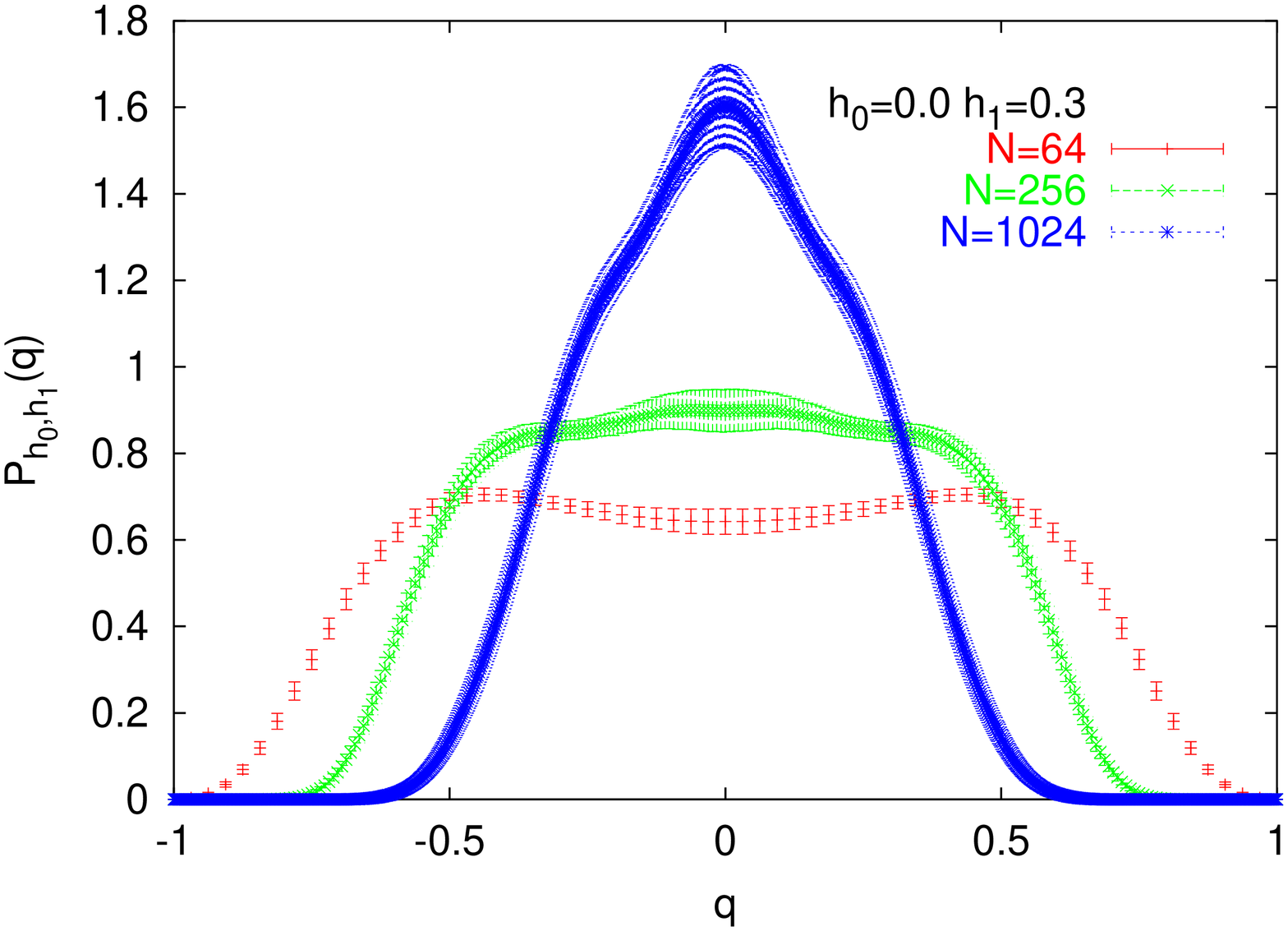,angle=270,width=8cm}
\caption{The probability distribution of the overlap $P_{h_0,h_1}(q)$ between 
replicas evolving at different magnetic field values, with $h_0=0.0$ and 
$h_1=0.1, 0.15, 0.2$ and 0.3 respectively, for the considered system sizes.}
\end{center}
\end{figure}

On the other
hand, to get a nearly Gaussian behavior we have to consider at least
$N=1024$ and $h_1$ values as large as 0.3, but the variance
is more than an order of magnitude larger than the one predicted 
by (\ref{pqh1h2}). Our data suggest that the support of $P_{0,h}$ 
shrinks to zero 
as $N$ grows, and  the chaotic $q\approx 0$ peak dominates more and more the 
distribution.

Moreover, we find that $A^{2n}(h_0,h_1)$ and $B^{2n}(h_0,h_1)$, which decrease for 
increasing sizes as soon as $h_1>0$, are in agreement with the expected 
scaling law \cite{Ri}, i.e. $\tilde{f}(N h_1^{8/3})$. We consider in particular
\be
B^2(h_0,h_1,T)=\frac{\overline{\langle \left ( q-\overline{\langle q 
\rangle}_{h_0,h_1} 
\right )^2 \rangle}_{h_0,h_1}}
{\overline{\langle \left ( q-\overline{\langle q 
\rangle}_{h_0,h_0} \right )^2 
\rangle}_{h_0,h_0}},
\ee
which is plotted in scaling form in [Fig. 4]. In the limit
$1/(Nh_1^{8/3})<<1$ the scaling function is  approaching the asymptotic
regime $\tilde{f}(Nh_1^{8/3}) \propto 1/(Nh_1^{8/3})$ in qualitative  
agreement 
with the first-order perturbative result (\ref{pqh1h2}). This shows that the 
asymptotic regime is indeed
approached in our data, and that we can safely deduce that
$\lim_{N\to\infty} B^2(0,h_1\neq 0,T)=0$.

\begin{figure}[htbp]
\begin{center}
\leavevmode
\epsfig{figure=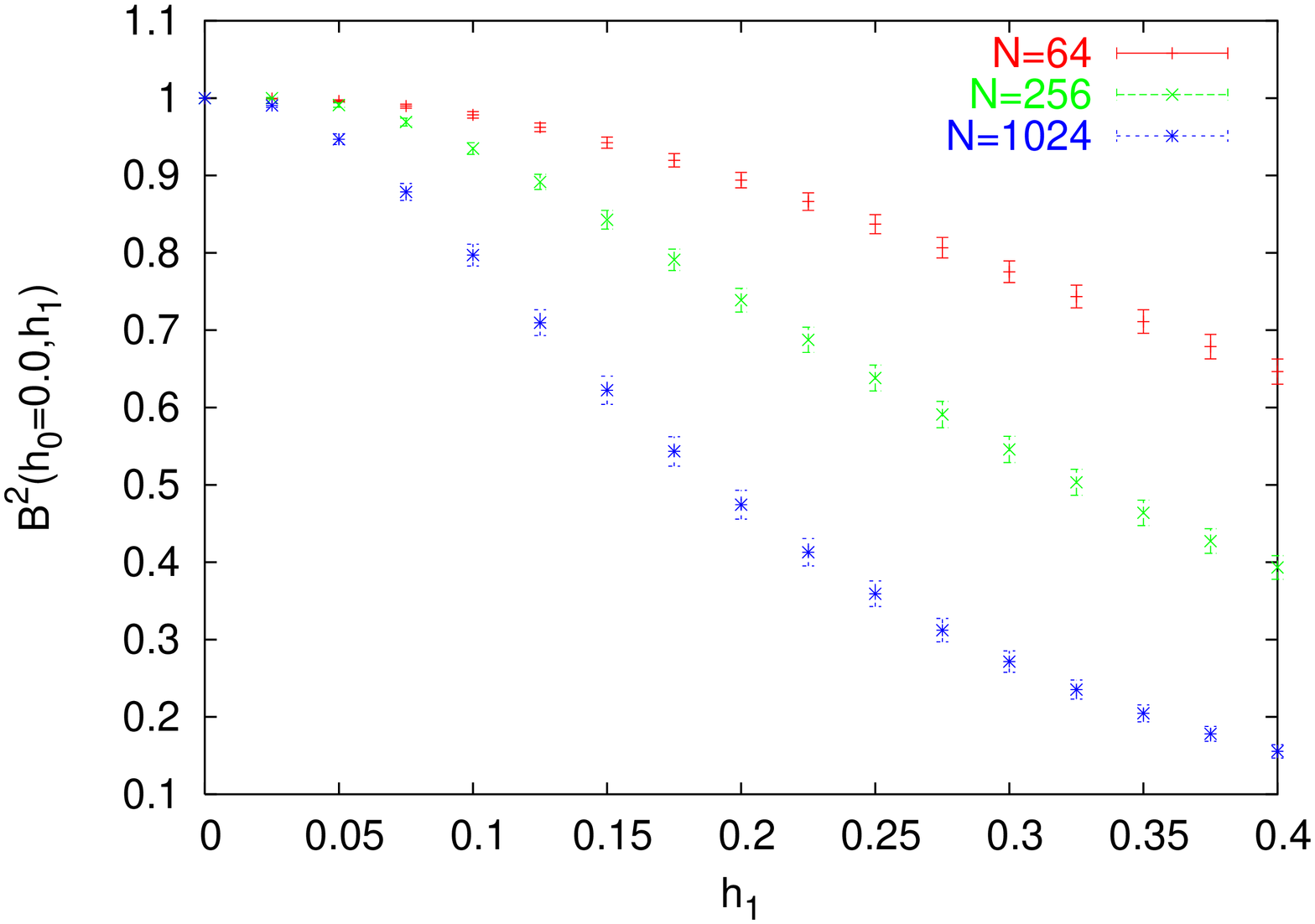,angle=270,width=8cm}
\epsfig{figure=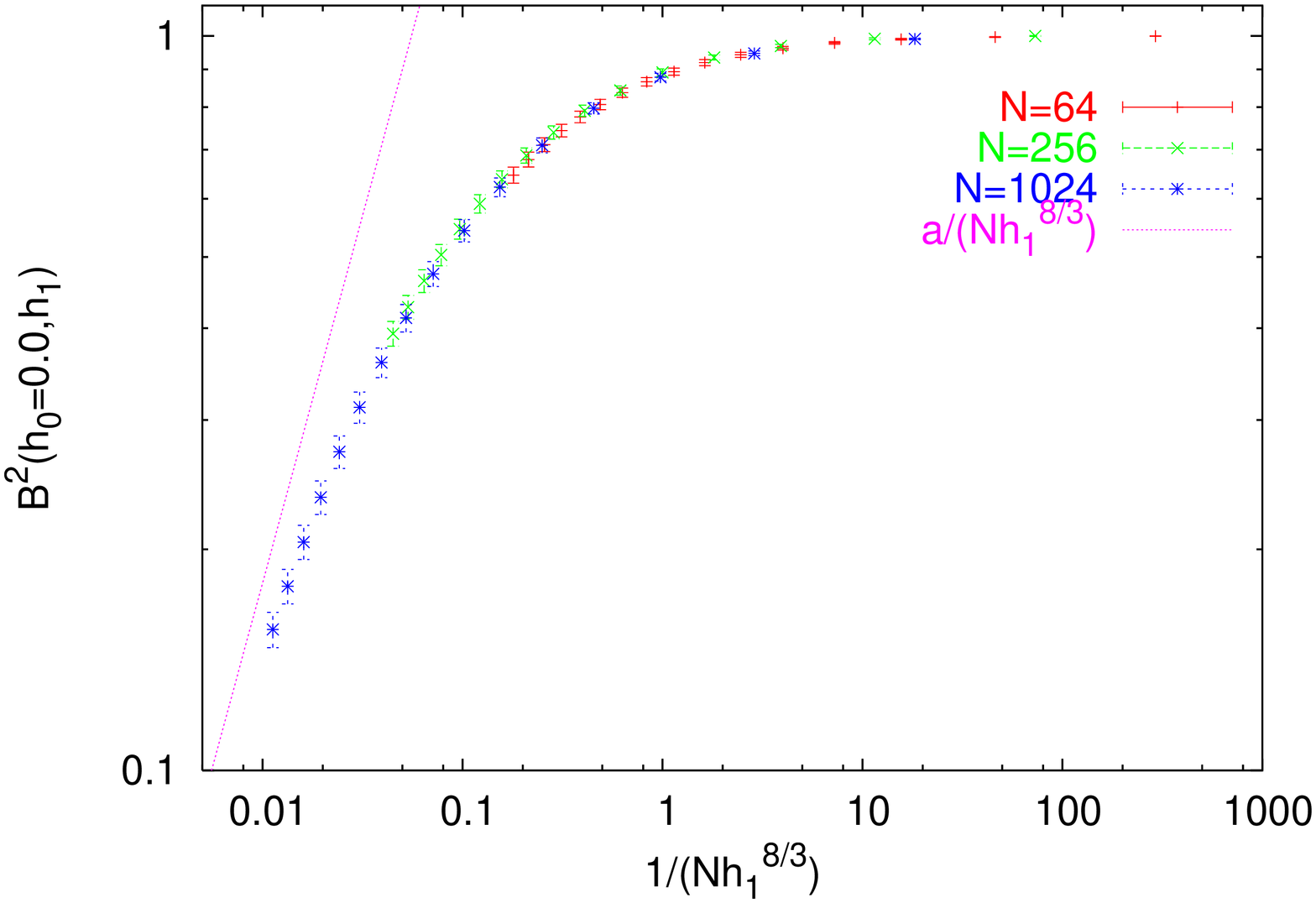,angle=270,width=8cm}
\epsfig{figure=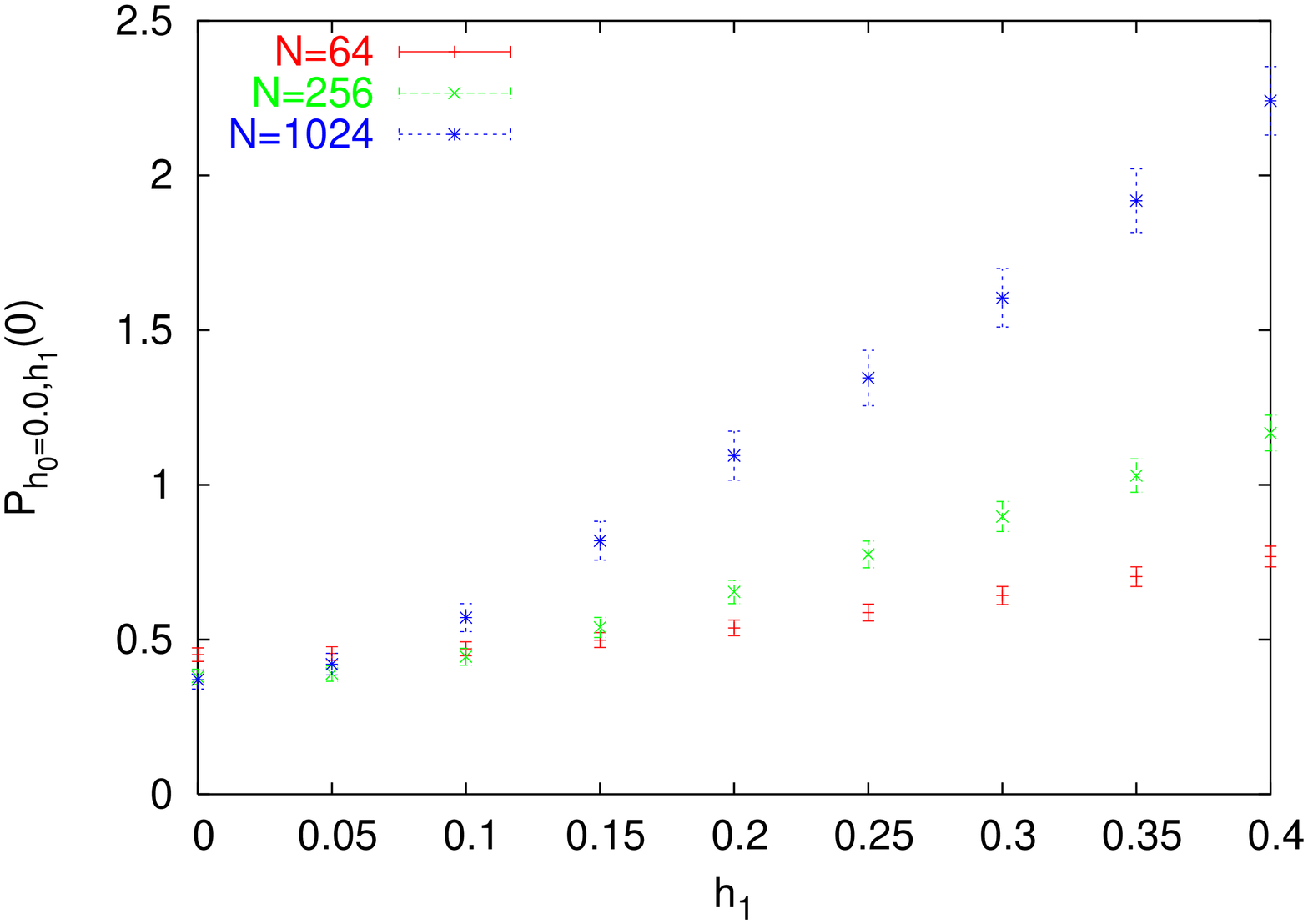,angle=270,width=8cm}
\epsfig{figure=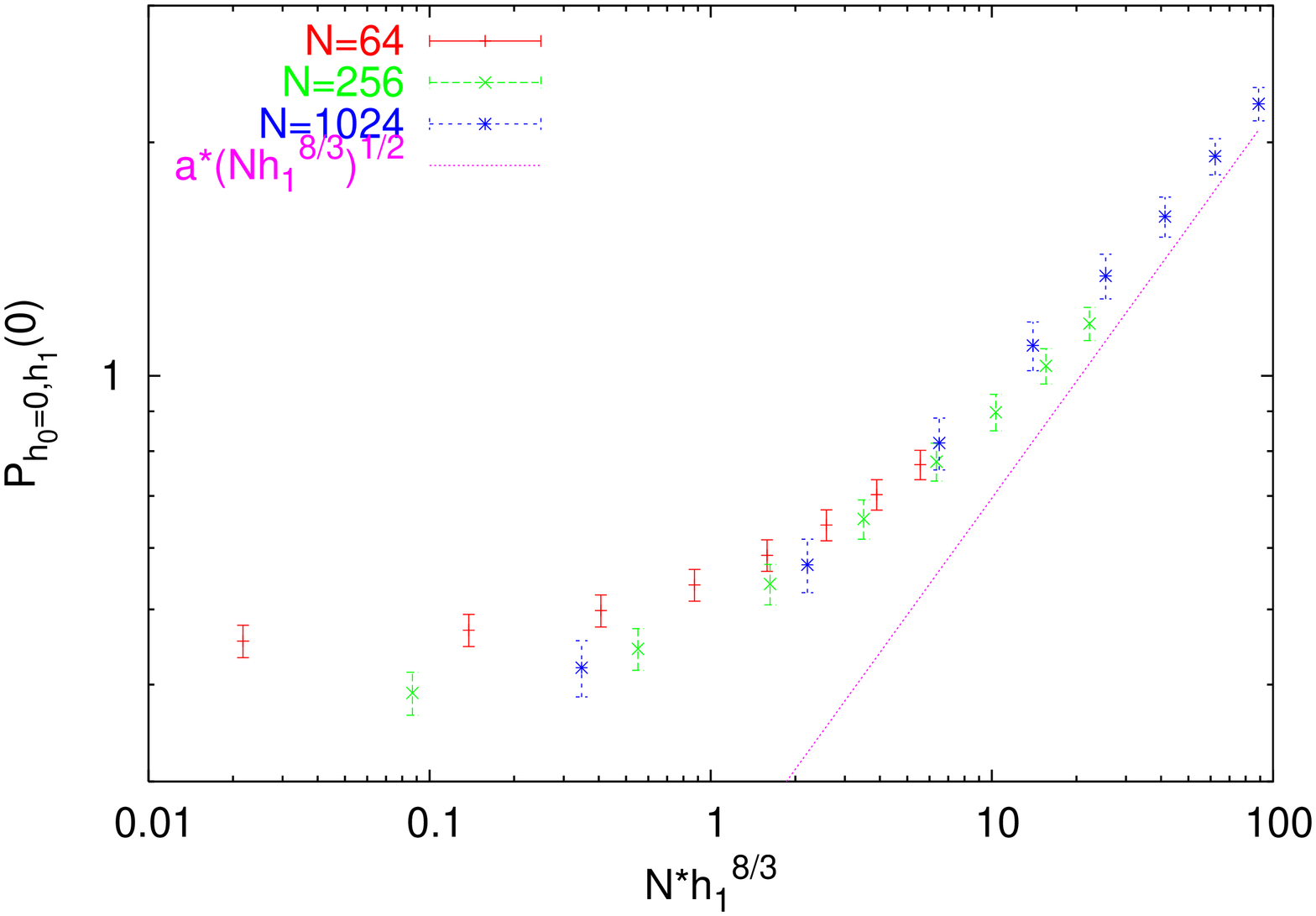,angle=270,width=8cm}
\caption{On top, the behavior of $B2(h_0,h_1)$ for $h_0=0.0$ as 
function of $h_1$ (left) and as function of the scaling variable
$1/(Nh_1^{8/3})$ compared with the asymptotic behavior $\propto
1/(Nh_1^{8/3})$ for $1/(Nh_1^{8/3}) <<1$ (log-log plot on the
right). On the bottom, the behavior of $P_{h_0,h_1}(0)$ for $h_0=0.0$
as function of $h_1$ (left) and as function of $Nh_1^{8/3}$ compared
with the asymptotic behavior $\propto \sqrt{Nh_1^{8/3}}$ for
$Nh_1^{8/3}>>1$ (log-log plot on the right).}
\end{center}
\end{figure}

\begin{figure}[htbp]
\begin{center}
\leavevmode
\epsfig{figure=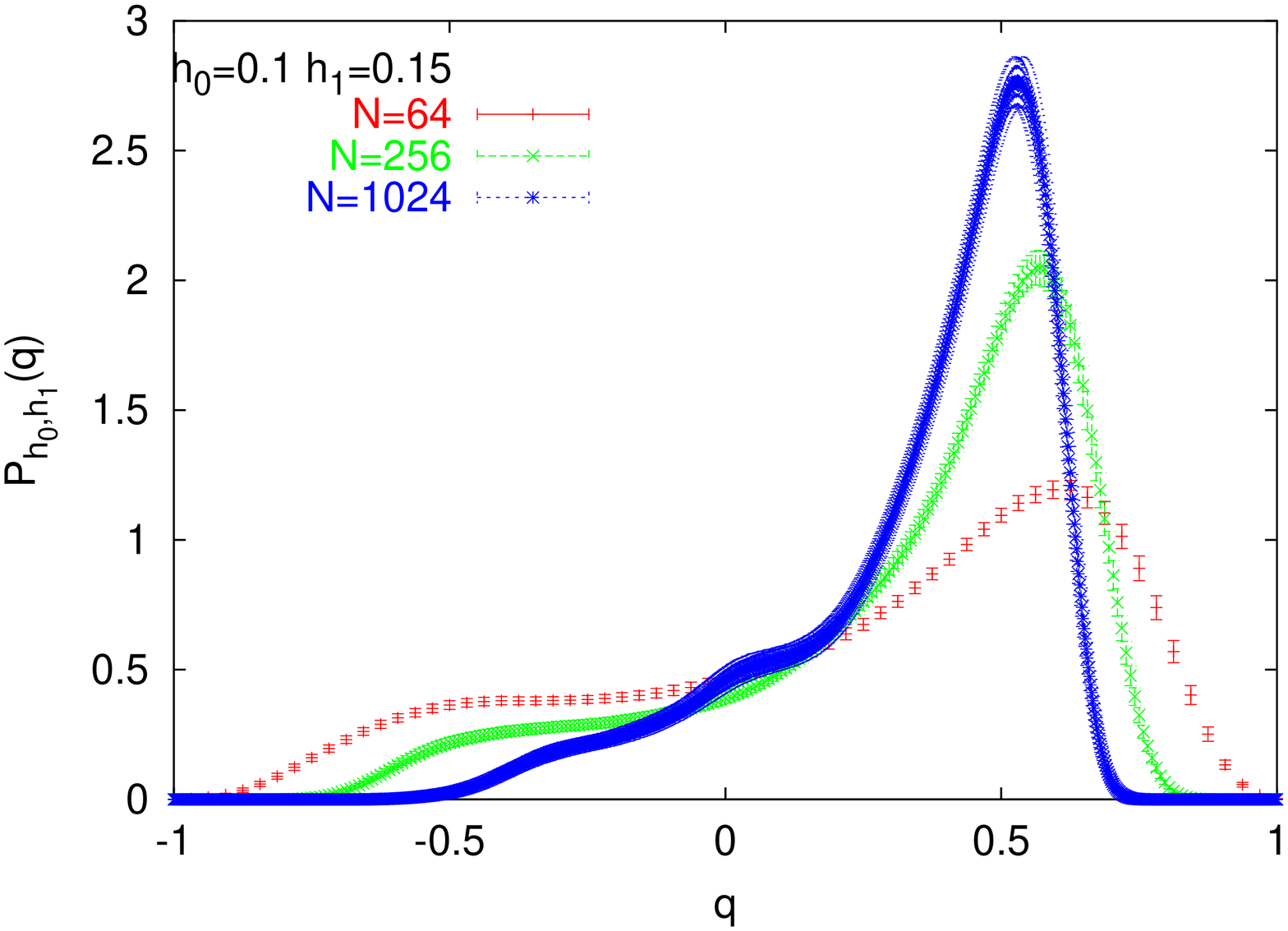,angle=270,width=8cm}
\epsfig{figure=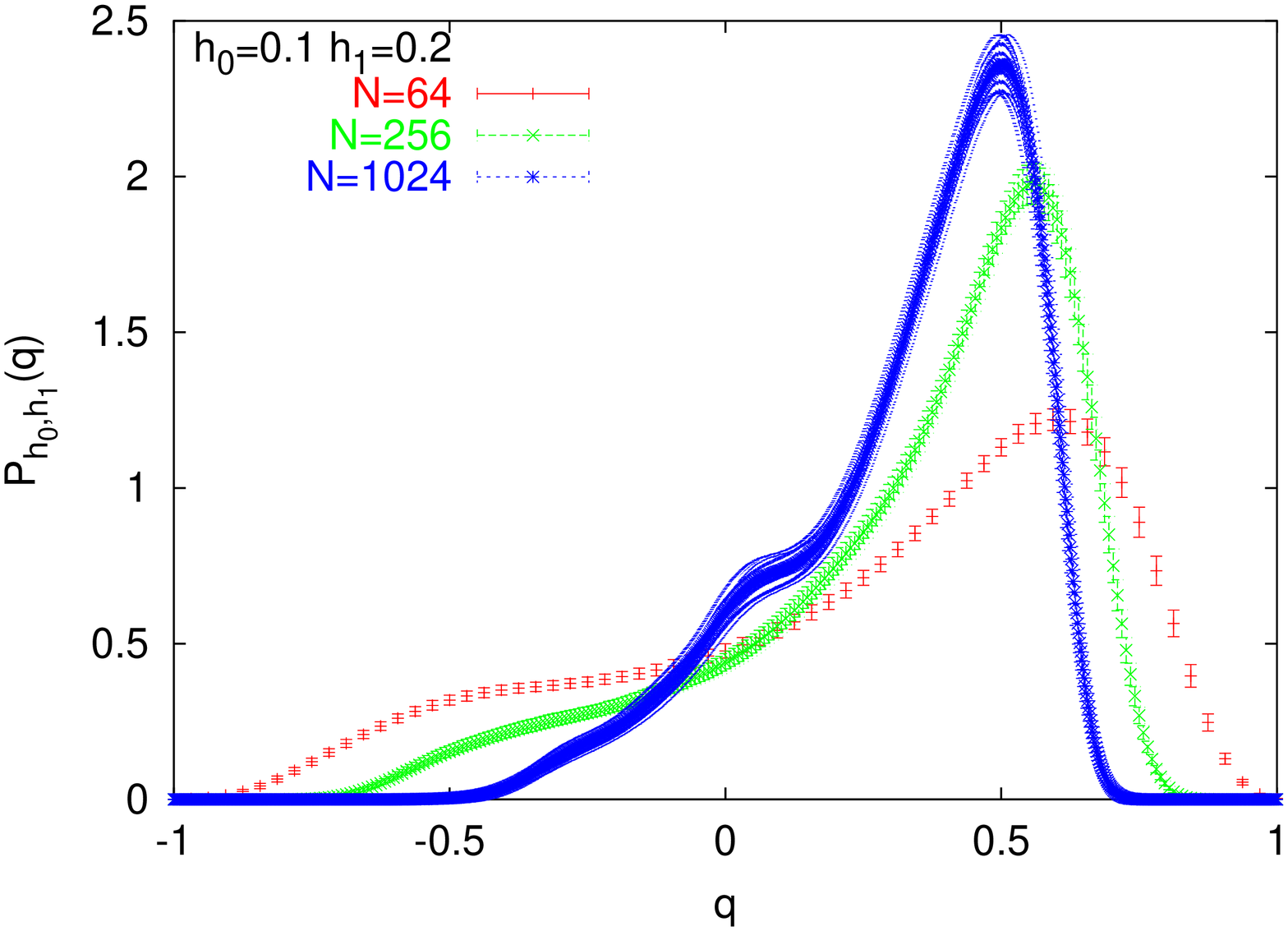,angle=270,width=8cm}
\epsfig{figure=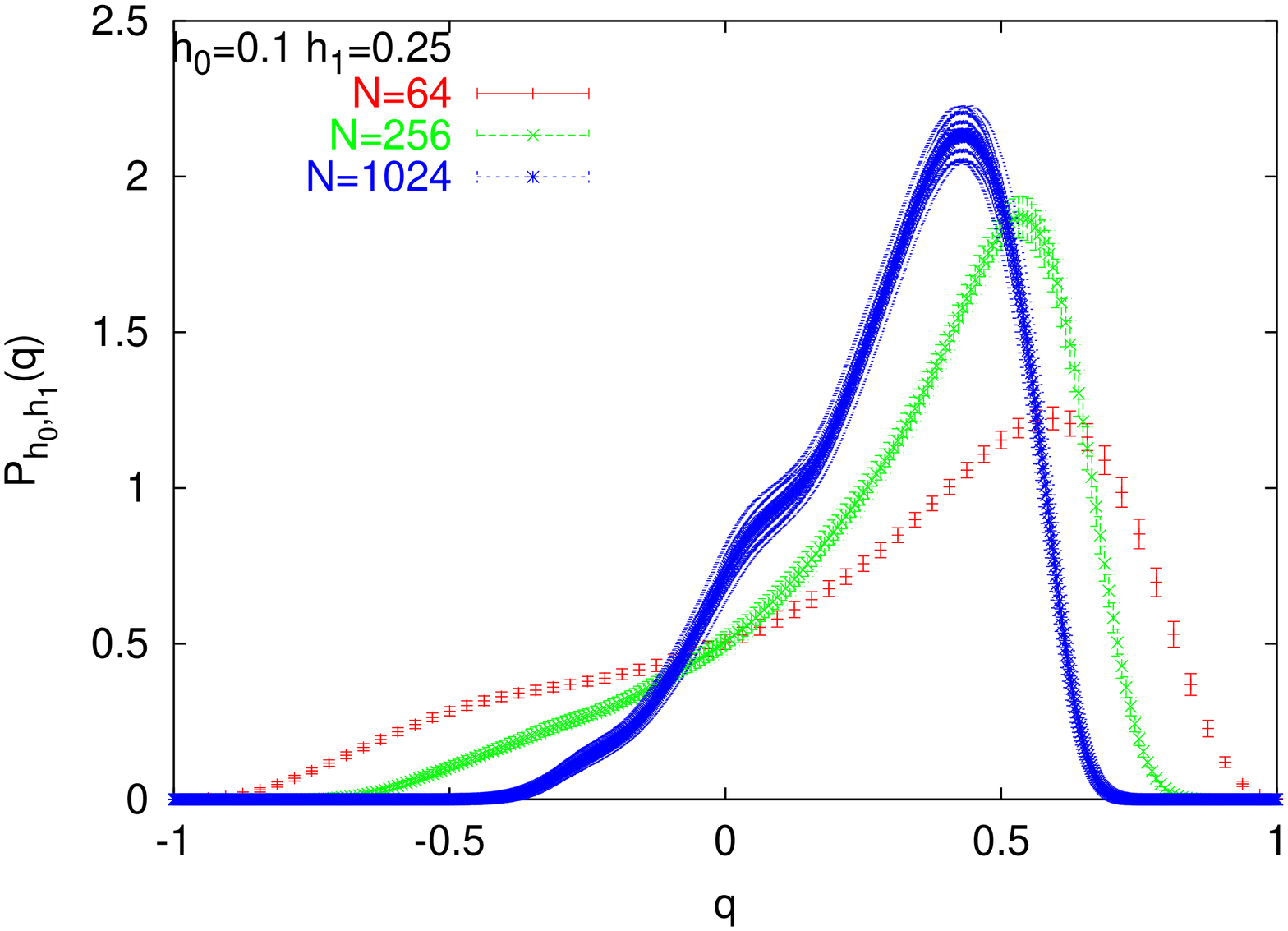,angle=270,width=8cm}
\epsfig{figure=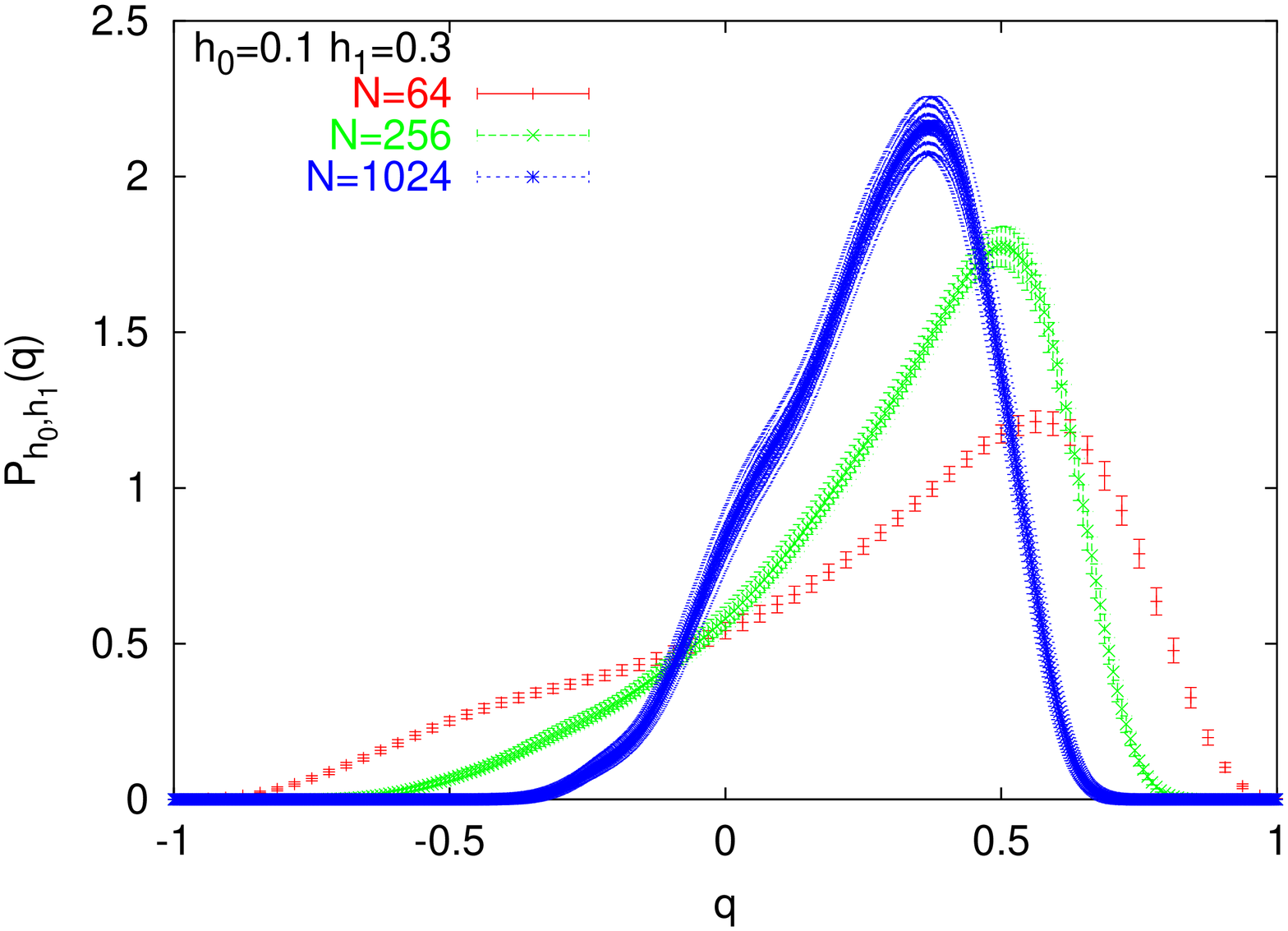,angle=270,width=8cm}
\caption{The probability distribution of the overlap $P_{h_0,h_1}(q)$ between 
two replicas evolving at different magnetic field values, with $h_0=0.1$ and 
$h_1=0.15, 0.2, 0.25$ and 0.3 respectively, for the considered system sizes.}
\end{center}
\end{figure}

\begin{figure}[htbp]
\begin{center}
\leavevmode
\epsfig{figure=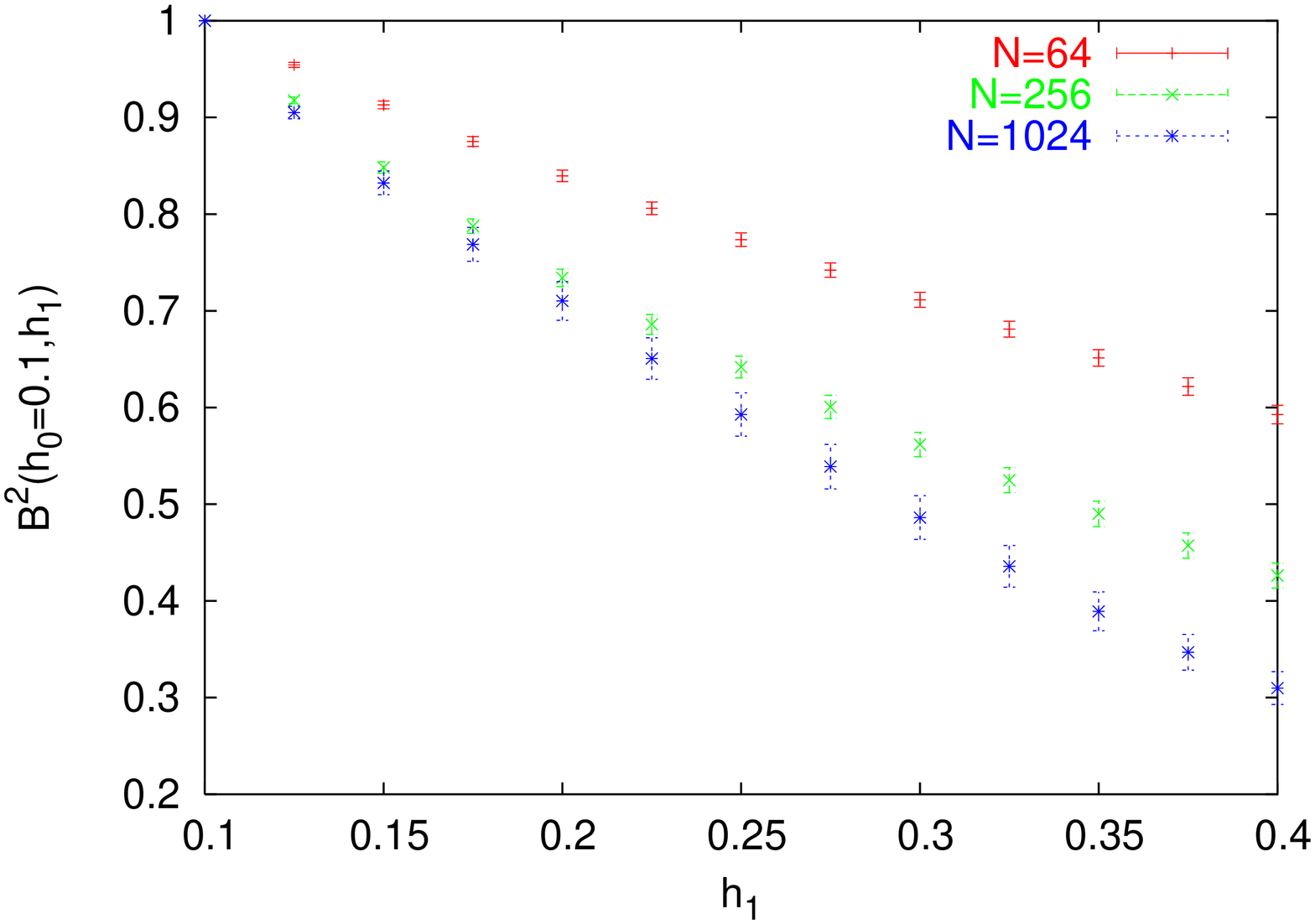,angle=270,width=8cm}
\epsfig{figure=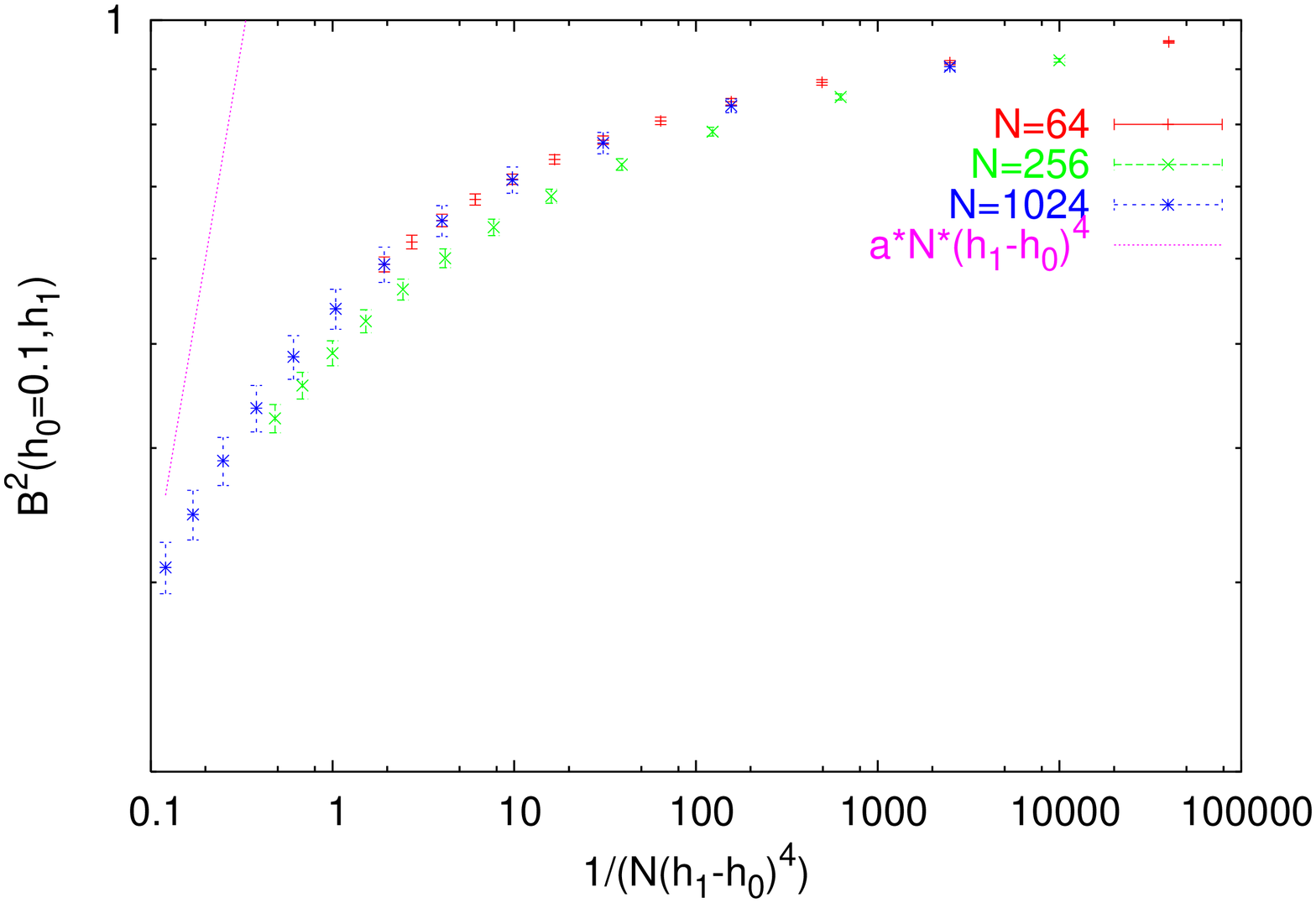,angle=270,width=8cm}
\caption{The behavior of $B2(h_0,h_1)$ for $h_0=0.1$ as function of $h_1$ 
(left) and as function of $1/(N(h_1-h_0)^{4})$ compared with the
asymptotic behavior $\propto 1/(N(h_1-h_0)^{4})$ (log-log plot on the
right).}
\end{center}
\end{figure}

We conclude the analysis of the $h_0=0$ case by looking at
$P_{h_0,h_1}(0)$ as a function of $h_1$. It increases when
considering increasing sizes (apart from the very small $h$ values,
where there are clearly strong finite size effects) and 
scales roughly as $\tilde{f}(N h_1^{8/3})$ (see [Fig. 4]) with
$\tilde{f}(Nh_1^{8/3})$ approaching the expected behavior $\propto
\sqrt{Nh_1^{8/3}}$ for $Nh_1^{8/3}>>1$. We also note that though we
have plotted data only for $h_1\le 0.4$ these scaling laws appear
satisfied also when including data corresponding to $h_1$ values on
the other side of the AT line.

Next we consider $h_0=0.1$, $h_1=0.15, 0.2, 0.25$ and $0.3$ (see [Fig. 5]). 
The $P_{h_0,h_1}(q)$ is still expected to approach a $\delta$-function in the 
thermodynamic limit, now centered in $q_m$
($q_m(h=0.1)=\qmin$  independent of $T$
from a recent analytical 
study \cite{CrRi1}).
However, we have 
already noted that the peak in $q_m$ is not evident in our data for (the usual) $P(q)$ and 
correspondingly there is no clear evidence for chaotic behavior in 
$P_{h_0\neq0,h_1}(q)$. Also for the largest size considered, i.e. $N=1024$, 
though a small 
second peak in $q \approx 0.05$ is appearing, the dominant contribution is 
still 
coming from the reminiscence of the peak in $q_{EA}$, whose mean value and 
height are slowly decreasing for increasing $h_1$. As a matter of fact,
when going to the other side of the AT line, i.e. $h_1 \ge 0.4$, it is this 
peak
that survives, becoming centered on a definitely lower $q$ value 
($\overline{\langle q \rangle}_{h_0=0.1,h_1=0.6}\simeq  0.18$ for $N=1024$, 
smaller than $q_m$). 

It is clear that one should look at larger $N$'s to get evidences for 
the expected Gaussian shape 
$\propto \exp({-(q-q_m)^2/2\sigma_{th}^2})$ (with 
$1/\sigma_{th}^2=\sqrt{2}Nh_0|h_1-h_0|$) in the spin glass phase. 
Therefore, it is not surprising that a quantity such as $B^2(h_0,h_1)$
does $not$ scale as a function of $N(h_1-h_0)$. In the case we are considering
of $h_0=0.1$ a form $B2\sim \tilde{f}(N(h_1-h_0)^{\alpha})$ still roughly 
works, with $\alpha \simeq 4$. Nevertheless, data presented in [Fig. 6] show 
that even for $N=1024$ and $(h_1-h_0) \simeq 0.3$ we are very far from an 
asymptotic regime $\tilde{f}(x) \propto 1/x$. When looking at larger $h_0$,
$B2$ definitely does not scale, for instance already at $h_0=0.2$ we get
$B2(N=1024)>B2(N=256)$ in the whole interval $0.2<h_1\le 0.4$.

\end{subsection}

\end{section}

\begin{section}{Conclusions}
\noindent
We performed numerical simulations of the SK model in a magnetic field
at temperature $T=0.6$ both in the glassy phase and above the AT
line. We used a modified version of the PT algorithm in which the system is
allowed to move between a chosen set of magnetic field values, an algorithm well
suited  for our purpose. We found that $N=1024$ is the
largest size one is able to efficiently thermalize with this method
and we argue that this is related to the appearance of 
magnetic field chaos  at this scale.

The function $P(q)$ shows strong finite size corrections
for $h>0$, with a long tail in the $q<0$ region that slowly disappears
for increasing sizes, whereas the peak corresponding to the thermodynamic
limit $\delta (q-q_m)$ is not yet visible.

Our main result is on the behavior of $P_{h_0,h_1}(q)$, which in the
case of $h_0=0.0$ shows evidence for chaos already at $h_1=0.15$ when
considering the still relatively small size $N=1024$. This is at
variance with the situation one finds when looking for temperature chaos
\cite{BiMa}, in agreement with the very recent analytical finding \cite{CrRi} 
that temperature chaos  is a much weaker effect. The appearance of the 
third peak in $q=0$ is accompanied by a shrinking of  the support of the distribution.

The expected scaling law \cite{Ri} is well satisfied and for large
$Nh_1^{8/3}$  $P_{h_0=0,h_1}(q)$ approaches a Gaussian with variance
$\propto 1/(Nh_1^{8/3})$, in qualitative agreement
with the result of a first order
perturbative computation \cite{FrNe}.

On the other hand, when looking at the chaotic behavior for $h_0\neq0$ we
found ourselves to be still very far from the expected asymptotic regime. This is
to be related to the presence of strong finite size effects observable
also on  $P(q)$ itself.

\end{section}

\begin{section}*{Acknowledgments}
\noindent
We acknowledge enlightening discussions with Andrea Crisanti, Cirano
De Dominicis, Silvio Franz, Enzo Marinari, Giorgio Parisi, Tommaso
Rizzo and Peter Young. B. C. is supported by a Marie Curie (CEE)
fellowship (contract MCFI 2001-00312). We thank  Andrea Crisanti and  Tommaso Rizzo for providing us 
with their unpublished results on the $h\neq 0$ SK model.
\end{section}

\end{document}